\documentclass[pdflatex,sn-mathphys-num]{sn-jnl}

\usepackage{graphicx}%
\usepackage{multirow}%
\usepackage{amsmath,amssymb,amsfonts}%
\usepackage{amsthm}%
\usepackage{mathrsfs}%
\usepackage[title]{appendix}%
\usepackage{xcolor}%
\usepackage{textcomp}%
\usepackage{manyfoot}
\usepackage{booktabs}%
\usepackage{algorithm}%
\usepackage{algorithmicx}%
\usepackage{algpseudocode}%
\usepackage{caption}
\usepackage{hyperref}
\usepackage{listings}%
\usepackage{tikz}
\usepackage{float}
\usepackage{url}

\theoremstyle{thmstyleone}%
\theoremstyle{thmstyletwo}%

\theoremstyle{thmstylethree}%

\newcommand{\subsubsubsection}[1]{
	\vspace{0.5em}
	\noindent\textbf{\textit{#1}}
	\par\nobreak\vspace{0.3em}
}

\begin{document}

\title[Robust Wave Origin Detection]{Robust Wave Origin Detection from Sensor Array Data via Randomized Hough Transform and Model Fitting}


\author[1]{\fnm{Sicheng} \sur{Fan}}\email{u12sf23@abdn.ac.uk}
\equalcont{These authors contributed equally to this work.}

\author[2]{\fnm{Jiayi} \sur{Lu}}\email{amylujy@gmail.com}
\equalcont{These authors contributed equally to this work.}

\author*[2]{\fnm{Xiaodan} \sur{Fan}}\email{xfan@cuhk.edu.hk}

\affil[1]{\orgdiv{Aberdeen Institute of Data Science and Artificial Intelligence}, \orgname{South China Normal University}, \orgaddress{\street{Taoyuan Dong Rd, Shishan Town,
Nanhai District}, \city{Foshan}, \state{Guangdong Province}, \postcode{528225}, \country{China}}}

\affil[2]{\orgdiv{Department of Statistics and Data Science}, \orgname{The Chinese University of Hong Kong}, \orgaddress{\street{Shatin}, \city{New Territory}, \state{Hong Kong SAR}, \country{China}}}


\abstract{The robustness of methods for characterizing the origin of wave-like propagating signals is often challenged by the heterogeneity of transmitting media and sensor network. Detecting an individual wave under such noisy conditions remains particularly difficult.
In this paper, we propose a robust framework to estimate wave origins from spatiotemporal data collected by a sensor array. First, a Butterworth filter is designed to extract relevant signals from each sensor channel. Next, a 3D randomized Hough transform is introduced to identify candidate signals originating from the same wave front. Finally, a wave propagation model is fitted to the identified signals using the least-squares method. By combining the iterative voting mechanism of the randomized Hough transform with least-squares optimization, our proposed method demonstrates superior robustness and computational efficiency in both simulation studies and real-world data analyses.}




\maketitle

\section{Introduction}\label{sec1}

The demand for study waves exists in many research fields, ranging from gastric slow waves to electromagnetic radiation. Therefore, the ability to detect and analyze waves has become essential. A common approach to detect waves is to deploy an array of sensors, such as microelectrode array
 for neuroscience \cite{example1-2025} and quantum dots focal‑plane arrays for infrared imaging \cite{example2-2026}. Compared with a single‑point detector, an array allows us to acquire signals from multiple spatial locations, thereby obtaining more accurate and diverse data about the waves, such as the orientation of the wave source, as well as its speed and propagation time, and the measurement of these parameters is what we focus on in this paper.

Usually, if the requirements of calculating the wave's propagation speed and its source are needed, we detect and analyze signal spikes. The Hough transform (HT) has become a powerful tool for handling such spike data. Originally designed for straight line detection in the 2D parameter space \cite{bib1}, HT has since been extended to circles and ellipses \cite{bib2} and, more recently, to planes and surfaces in the 3D parameter space \cite{bib3}. These developments provide a solid theoretical foundation for the identification of geometric models in many application domains.

In the context of wave detection, most existing studies on the processing detector array's data are operated in two‑dimensional parameter spaces since most units on the detector array are arranged in one-dimensional (linear) form \cite{bib4}. But for cases where detector arrays are used—such as Membrane Electrode Assembly (MEA) for detecting bio-electricity signals or Quantum dot focal plane array (FPA) for detecting infrared radiation, the number of parameter space expands from two-dimensional to three-dimensional, adding time dimension. For this reason, no matter the complexity or time of the calculation, it will increase. Thus, an efficient algorithm for processing three-dimensional data is required.

With traditional processing methods, 3D parameter space data could be hard to process. The Hough transform has indeed been widely applied in 3D parameter processing. For example, recognizing 3D models' curves that can be parameterized by mathematical means \cite{bib6}, and for another, processing the 3D point cloud maps obtained through scanning methods such as lidar, for mathematical modeling and analysis \cite{bib7}, which may be the most common application of 3D Hough transform. However, few cases are found in the 3D Hough transform field for detecting waves, which is also an important analytical point in many disciplines.

To fill this gap, we propose a Randomized 3D Hough transform method for efficient plane detection from detector array data. After identifying the wave plane, we fit a linear model to obtain the wave source's orientation, propagation speed, and recording time. The proposed method achieves comparable or superior accuracy while requiring significantly less computational time than conventional deterministic approaches.

The remainder of this paper is organized as follows. Section~\ref{sec2} describes the whole methodology of the paper; Section~\ref{sec3} presents experimental results on synthetic and real datasets; finally, Section~\ref{sec4} concludes with future research directions.

\section{Method}\label{sec2}
\begin{figure}[ht]
	\centering
	\includegraphics[scale=0.34]{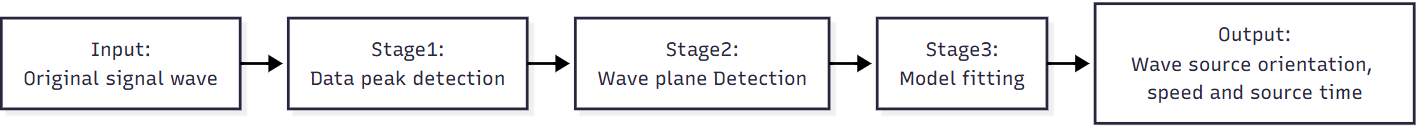}
	\captionsetup{justification=centering}
	\caption{A multistage procedure for analyzing wave signal spike data.}
	\label{fig:Process_Flow}
\end{figure}

In this section, a method will be introduced to detect the wave plane in parameter space. The whole process can be divided mainly into the following few steps in the flow chart \ref{fig:Process_Flow}.

As can be discerned from the Fig~\ref{fig:Process_Flow}, the detection process mainly consists of three stages. The first stage is for preprocessing the data, includes receiving wave data, conducting noise reduction processing and detecting the peak value of the signal. The second is for detecting wave propagation planes by Randomize Hough transform. The last is to fit the output model using the alternating variable method. In the whole procedure, the initial input is a data set with spike locations and times reflecting the characteristic of wave. Between different stages, the output of one stage is the input of the stage which follows. The final outputs are locations or directions of wave sources. Each stage will be illustrated one by one in the following subsections.

\subsection{Data Peak Extraction}
When using a detector array, for every unit, what we actually received are wave signals. Usually, for these signals, factors that are unfavorable for analysis, such as noise, are unavoidable. Therefore, to guarantee data accuracy and to facilitate subsequent research, we need to denoise the signal and extract the peaks of the signal.

In this paper, the Butterworth high-pass filter is applied to the input time series to obtain \(y*\), and locate the first candidate point \(t_0\) through global peak detection. Then, use the $99.95\%$ quantile as a dynamic threshold to mark spikes when the amplitude condition is met. At the same time, exclude the time neighborhood range of the detected spikes \(G\) in each iteration, and continue to search for the next largest peak until the threshold condition is not met. This design removes low-frequency interference through high-pass filtering, uses a high percentile threshold to screen significant spikes, and combines a neighborhood exclusion mechanism to ensure the temporal accuracy of spike detection. Algorithm \ref{alg:butterworth} offers an implementation of the procedure mentioned above.

\begin{algorithm}
	\caption{Butterworth algorithm for finding spike activation time}
	\label{alg:butterworth}
	\begin{algorithmic}[1]
		\State Given input time series $y: [0, T] \to \mathbb{R}$
		\State Apply the Butterworth high-pass filter to get a new time series $y^*: [0, T] \to \mathbb{R}$
		\State Find $t_0 = \arg\max_{t\in [0,T]}(y^*(t))$
		\State Set $G$ = minimum time gap between spikes, $i = 0$
		\While{$y^*(t_i) > \textrm{99.95 percentile of } \{y^*(t)\}_{t\in [0,T]}$}
		\State Mark $t_i$ as a spike
		\State \(t_{i+1} =
		\arg\max_{
			t\in [0,T]\backslash \cup_{k=0}^i [t_k - G, t_k+G]
		}(y^*(t))
		\)
		\State $i=i+1$
		\EndWhile
	\end{algorithmic}
\end{algorithm}

Then the result will no longer be a continuous signal, but a set representing the location of the signal spikes $(x, y)$ and their occurrence time $t$, which is $(x, y, t)$. This will greatly facilitate subsequent transformation and other steps.

\subsection{Randomized Hough transform}
After extracting the spike activation time, we need to separate the signal spikes into
multiple propagating waves. Since the wave passes through the detector array at a constant homogeneous speed in all directions, the recorded signal spikes form a plane in the 3D space of $(x, y, t)$, where $(x, y)$ is the coordinate of the detected spike, and $t$ is the time when the spike occurs. To find such planes, we apply the Randomized Hough transform-a Randomized alternative to the standard Hough
transform proposed by Borrmann et al \cite{bib8}.

Compared with the Randomized Hough transform, in order to adapt to the wave detection task, we applied some optimizations. In addition to setting a threshold on the accumulator, to fit the scenarios where detector array is used, a threshold of total point number on detected plane is programmed, which is decided by the number of detector unit in the array(in here, the threshold is two-thirds of the detector unit). In addition, there is a dedicated for loop to handle the points that were not initially classified. It attempts to group them back into the discovered planes based on "proximity", which is usually the distance between the point and plane. The main steps are summarized in Algorithm \ref{alg:randomhough}

\begin{algorithm}
	\caption{Randomize Hough transform for wave detection}
	\label{alg:randomhough}
	\begin{algorithmic}
		\State Given a set of spikes $\{(x_i, y_i, t_i)\}_{i=1}^n$
		\While{}
			\State Run randomized Hough transform until a bucket gets 8 votes
			\State Check every remaining point to see if it is in the plane
			spanned by this bucket
			\If{Number of points on the plane $> \frac{2}{3}$ (Number of detectors)}
				\State Mark the points on this plane as a cluster and remove all
				these points
			\Else
				\State Exit while loop
			\EndIf
		\EndWhile
			\For{Each point $P$ that's not marked to any plane}
			\If{$P$ is sufficiently close to a plane found above}
				\State Add $P$ to that plane.
			\EndIf
			\EndFor{}
		\State The unclassified points will be regarded as noise.
	\end{algorithmic}
\end{algorithm}

\subsection{Modeling and Fitting} \label{sec:gridoptim}
In the previous section, we divided the original spike data into several planar clusters through the randomization of the Hough transform, with each cluster corresponding to a separate wavefront.

The task of this chapter is to perform parameter fitting for each detected wavefront. For this, we present two common geometric assumptions: If the wave source is close to the detector, the wavefront can be regarded as circular; if the wave source is far away, the wavefront can be considered as a straight line locally. For these two types of model, we will respectively provide the arrival-time formulas and construct the loss function based on them.

When it comes to fitting process, for the circular model, due to the high coupling degree among its parameters, we adopt the alternating minimization strategy: first fix the speed to determine the source point, then fix the source point to determine the speed; repeat the iteration until convergence. Then, for linear models, the problem can be transformed into a standard least-squares regression, and a closed-form solution can be directly obtained, thereby quickly obtaining the wavefront normal vector and propagation rate.

Through these fitting procedures, we are able to recover many descriptions of a wave, such as the location of the wave source, its propagation speed, the excitation time, etc.

\subsubsection{Wave Propagation Models}
\subsubsubsection{Near‑Field Circular Wavefront Model}
Let us first focus on the circular model, which is employed when wave sources are close to the detector or in the detector array. We assume that the wave source is located at $(x_0, y_0)$, and meanwhile, assume that the propagation of the signal has the same speed $v$ in all directions. As for the independent measurement error of the recording signal $e_k$ for each node of the detector array $k$, we have the following:
\[
e_k \sim N(0, \sigma^2).
\]
Here, $N(0, \sigma^2)$ denotes the normal distribution with mean $0$ and variance $\sigma^2$. Therefore, the recording time of the signal at the coordinate $(x_k, y_k)$ can be expressed as
\begin{equation}\label{eqn:arrival_time_cone}
	t_k = \sqrt{(x_k-x_0)^2+(y_k-y_0)^2} / v + e_k.
\end{equation}

\subsubsubsection{Far‑Field Linear Wavefront model}
When sources are far from the detector, which means that the case when estimated $(x_0,y_0)$ may not be informative, we use a linear model. Consequently, we should instead formulate the signal source as linear with the form:
\[
ax+by+c=0,
\]
where $\vec{n}=(a,b)$ denotes the normal vector of this line. Combined with the traveling distance for the linear signal wave to arrive at the coordinates $(x_k, y_k)$ on the detector array:
\[
s_k = \frac{ax_k+by_k+c}{\sqrt{a^2+b^2}},
\]
The corresponding recording time of the signal will be:
\begin{eqnarray}\label{eqn:arrival_time_line}
	t_k &=& \frac{s_k}{v} + e_k \nonumber \\
	&=& \frac{ax_k+by_k+c}{\sqrt{a^2+b^2} \cdot v} + e_k.
\end{eqnarray}

For this linear wavefront model, the signs of $a$ and $b$ represent the propagating direction of the signal. The relation is shown in Table \ref{tab:relation}.
\begin{table}[htbp]
	\centering
	\caption{Relation between the signs of $a,b$ and the propagating direction of the signal}
	\label{tab:relation}
	\begin{tabular}{*{4}{c}}
		\toprule
		\multicolumn{2}{c}{\textbf{Sign}} & \multicolumn{2}{c}{\textbf{Direction}} \\
		\cmidrule(lr){1-2} \cmidrule(lr){3-4}
		$a$ & $b$ & \textbf{vertical} & \textbf{horizontal} \\
		\midrule
		$+$ & $+$ & down & right \\
		$-$ & $+$ & up & right \\
		$+$ & $-$ & down & left \\
		$-$ & $-$ & up & left \\
		\bottomrule
	\end{tabular}
\end{table}

The three unknowns $(a,b,c)$ are equivalent (up to a scalar factor) to the direction and offset of the wavefront; we will estimate them by fitting the least squares in the following content.

\subsubsection{Fitting Process}
\subsubsubsection{Circular wavefront model}
\newcommand{\closs}{\ell_{\textrm{circular}}}
To fit the circular wavefront model introduced above, which is formula~\ref{eqn:arrival_time_cone}, loss function $\closs$ is defined:
\begin{equation}
	  \label{eqn:circularloss}
	\closs(x_0, y_0, v, t_0)
	= \sum_{i=1}^n \left(
	\frac{\sqrt{(x_i-x_0)^2+(y_i-y_0)^2}}{v} - (t_i - t_0)
	\right)^2
\end{equation}

where $\{(x_i, y_i, t_i)\}_{i=1}^n$ is a set of spikes classified in the same
plane generated after the Randomize Hough transform. For other parameters, $t_0$ is the time when excitation occurs, $(x_0, y_0)$ is the exact location of the wave source, and $v_0$ is the wave propagation velocity. Our goal is to minimize $\closs$ with respect to $(x_0, y_0, v, t_0)$. Direct calculation of the four parameters can be difficult and complex since its parameter space has up to 4 dimensions, so instead we employ an alternating variable method summarized in Algorithm \ref{alg:alternating} to reduce the complexity of calculation. We will elaborate on how $\min_{x_0, y_0} \closs(x_0, y_0, v^{(i)}, t_0^{(i)})$ and $\min_{v, t_0} \closs(x_0^{(i+1)}, y_0^{(i+1)}, v, t_0)$ are computed below.

\begin{algorithm}
	\caption{Minimizing loss function for circular wavefront model}
	\label{alg:alternating}
	\begin{algorithmic}
		\State Given a set of spikes $\{(x_i, y_i, t_i)\}_{i=1}^n$
		\State Given initial guess $t_0^{(0)}$, $v^{(0)}$
		\State $i = 0$
		\While{}
			\State $x_0^{(i+1)}, y_0^{(i+1)}
			= \arg\min_{x_0, y_0} \closs(x_0, y_0, v^{(i)}, t_0^{(i)})$
			\State $v^{(i+1)}, t_0^{(i+1)}
			= \arg\min_{v, t_0} \closs(x_0^{(i+1)}, y_0^{(i+1)}, v, t_0)$
			\State $\Theta = (x_0^{(i)}, y_0^{(i)}, v^{(i)}, t_0^{(i)})$
			\State $\hat{\Theta} = (x_0^{(i+1)}, y_0^{(i+1)}, v^{(i+1)}, t_0^{(i+1)})$
			\If{$\| \Theta - \hat{\Theta} \| / \| \Theta \| < 10^{-6}$}
        		\State Exit while loop
			\EndIf
			\State $i=i+1$
		\EndWhile
	\end{algorithmic}
\end{algorithm}

\noindent
\\\textbf{Minimizing $\closs$ with respect to $x_0, y_0$}

To optimize $x_0, y_0$, we design an iterative tabulation method.
Suppose that we want to find the point $(a, b)$ where a function
$g: [L_a, U_a]\times [L_b, U_b] \to \mathbb{R}$ attains minimum. We
divide the search space into a 4-by-4 lattice with
\[
a_0 = L_a,
a_1 = \frac{2}{3}L_a + \frac{1}{3}U_a,
a_2 = \frac{1}{3}L_a + \frac{2}{3}U_a,
a_3 = U_a,
\]
\[
b_0 = L_b,
b_1 = \frac{2}{3}L_b + \frac{1}{3}U_b,
b_2 = \frac{1}{3}L_b + \frac{2}{3}U_b,
b_3 = U_b.
\]
And we compute
\[
g_{i,j} = g(a_i, b_j) \qquad (i = 0,1,2,3 \quad j = 0,1,2,3)
\]
Among all the $g_{i,j}$, we find $i^*, j^*$ that gives the minimum value.
Then define $\hat{i}, \hat{j}$ as follows
\[
\hat{i} =
\begin{cases}
	1  &(i^* = 0) \\
	i^*  &(i^* = 1, 2) \\
	2  &(i^* = 3)
\end{cases}
\qquad
\hat{j} =
\begin{cases}
	1  &(j^* = 0) \\
	i^*  &(j^* = 1, 2) \\
	2  &(j^* = 3)
\end{cases}
\]
In the next iteration, the search space is reduced from
$[L_a, U_a]\times [L_b, U_b]$ to
$[a_{\hat{i} - 1}, a_{\hat{i} + 1}] \times [b_{\hat{i} - 1}, b_{\hat{i} + 1}]$.
Intuitively, this is like zooming into a smaller region of the search space where the sample points indicate the function value is smaller there. We continue to divide the smaller search space into a four-by-4 lattice compute function value for $g$. We zoom in to the region that gives a smaller function value again. This iterative process continues until the size of the search space is smaller than our error tolerance. Then we obtain the final result of $a_0$ and $b_0$.

The whole process will be applied to optimize $x_0, y_0$ replacing $a$ with $x$ and $b$ as $y$ in the above process, and for a wave plane data, $(x_0, y_0)$ will be the coordinate of the corresponding wave source. To optimize $u_0$ and $t_0$, $x_0, y_0$ will remain constant, and we apply the following process.

\noindent
\\\textbf{Minimizing $\closs$ with respect to $v, t_0$}

Let $u = \frac{1}{v}$. When $x_0, y_0$ are kept constant,
\[
h(u, t_0) = \closs(x_0, y_0, 1/u, t_0)
= \sum_{i=1}^n \left(
u\sqrt{(x_i-x_0)^2+(y_i-y_0)^2} - (t_i - t_0)
\right)^2
\]
is a quadratic function in $u, t_0$. Thus, we can view this as a linear regression problem.
\[
t_i \sim u\sqrt{(x_i-x_0)^2+(y_i-y_0)^2} + t_0,
\]
which is a linear model with $u$ as slope and $t_0$ as intercept.
Optimal $u^*$ and $t_0^*$ can be found by fitting this linear model using the least squares method.

\subsubsubsection{Linear Wavefront Model}
\label{sec:linearfit}
\newcommand{\ta}{\tilde{a}}
\newcommand{\tb}{\tilde{b}}
\newcommand{\tc}{\tilde{c}}
\newcommand{\lloss}{\ell_{\textrm{linear}}}

To fit the linear wavefront model introduced in the Introduction section, we define the loss function by rewriting the function \eqref{eqn:arrival_time_line} as:
\begin{align}
	\lloss(\ta, \tb, \tc)
	&= \sum_{i=1}^n (\ta x_i + \tb y_i + \tc - t_i)^2
	\label{eqn:linearloss} \\
	&= \sum_{i=1}^n \left( \frac{ax_i + by_i + c}{v \sqrt{a^2 + b^2}} - (t_i - t_0) \right)^2
	\nonumber
\end{align}
with reparameterized coefficients:
\[
\ta = \frac{a}{v \sqrt{a^2 + b^2}}, \quad
\tb = \frac{b}{v \sqrt{a^2 + b^2}}, \quad
\tc = \frac{c}{v \sqrt{a^2 + b^2}} + t_0
\]
Given spike data $\{(x_i, y_i, t_i)\}_{i=1}^n$, we solve the following:
\begin{equation}
	\label{eqn:minlinearmodel}
	(\ta^*, \tb^*, \tc^*) = \arg\min_{\ta, \tb, \tc} \lloss(\ta, \tb, \tc)
\end{equation}

The closed-form solution via least squares is:
\begin{align}
	\begin{bmatrix}
		\ta^* \\
		\tb^* \\
		\tc^*
	\end{bmatrix}
	&= (\mathbf{X}^T \mathbf{X})^{-1} \mathbf{X}^T \mathbf{T}
	\label{eq:matrix_solution} \\
	\mathbf{X} &= \begin{bmatrix}
		x_1 & y_1 & 1 \\
		x_2 & y_2 & 1 \\
		\vdots & \vdots & \vdots \\
		x_n & y_n & 1
	\end{bmatrix},
	\quad
	\mathbf{T} = \begin{bmatrix}
		t_1 \\
		t_2 \\
		\vdots \\
		t_n
	\end{bmatrix}
	\label{eq:design_matrix}
\end{align}

The wave propagation speed is derived as follows:
\[
v^* = \frac{1}{\sqrt{(\ta^*)^2 + (\tb^*)^2}}
\]
With the results above, we can determine the direction of wave propagation and its speed.

Since $(\ta, \tb) = \lambda (a, b)$ where $\lambda = \frac{1}{v \sqrt{a^2 + b^2}}$ is a scalar, the direction vectors $(\ta, \tb)$ and $(a, b)$ are equivalent. We therefore use $(\ta^*, \tb^*)$ to represent the wave propagation direction in a subsequent analysis.

\section{Experiments}\label{sec3}

\subsection{Simulation Study based on Synthesized Data}
In this part, we will use synthetic data to demonstrate the performance of the randomized Hough transform algorithm (Algorithm \ref{alg:randomhough}) as well as the model fitting method(Algorithm \ref{alg:alternating}). Data generated under the linear wavefront model will be used to conduct the Hough transform, while the model-fitting method is applied to those simulated by the circular wavefront model. Performance evaluation on wave detection and parameter estimation will be reported.

\subsubsection{Wave detection using Hough transform}
\paragraph{Data synthesis from linear wavefront model}
In order to rigorously evaluate the performance and robustness of the proposed generalized Randomized Hough Transform (RHT) algorithm, numerical simulation on controlled data is indispensable. Simulation allows for a precise understanding of the algorithm's behavior under known conditions, which is a crucial step before applying it to complex and often noisy experimental recordings.

Here, we have designed a simulation model that is highly similar to the typical MEA experimental setup. We considered an MEA grid in the Euclidean space arranged in an 8-by-8 pattern, which is a common configuration that can provide sufficient data points for analysis and also ensure efficient computation during repeated tests. We design that signals are propagating on this MEA grid in such a manner:

\begin{itemize}
	\item The wave source is in the corner of the grid. Suppose that its coordinate is $(1,1)$.
	\item The duration of the whole simulation study is $t_{\max}$. The MEA stops recording signals when $t_{\max}$ is reached.
	\item The signal spike is propagating in the positive directions of the $x$ and $y$ axes simultaneously, where $x$ denotes the row and $y$ denotes the column. Once the signal reaches the next row (i.e. $x+1$), it starts to propagate exclusively in the y direction.
	\item The travel time between two adjacent electrodes in the $x$ direction follows a normal distribution with mean $\mu_x$ and variance $\sigma^2_x$, while in the $y$ direction it follows a normal distribution with mean $\mu_y$ and variance $\sigma^2_y$.
	\item For each electrode, there is a probability $p$ that the signal spike cannot be detected.
	\item Multiple signal waves from the same source can propagate in the MEA one after another. The time gap between two excitation times of the signal follows an exponential distribution with mean $\lambda_s$. The first signal wave is excited at time $t_0$.
	\item Noise spikes are also recorded by the MEA system, whose locations are randomly selected from all 64 electrodes. The time gap between two noise spikes follows an exponential distribution with mean $\lambda_n$.
\end{itemize}

\begin{figure}[!ht]
	\centering
	\includegraphics[width=\textwidth]{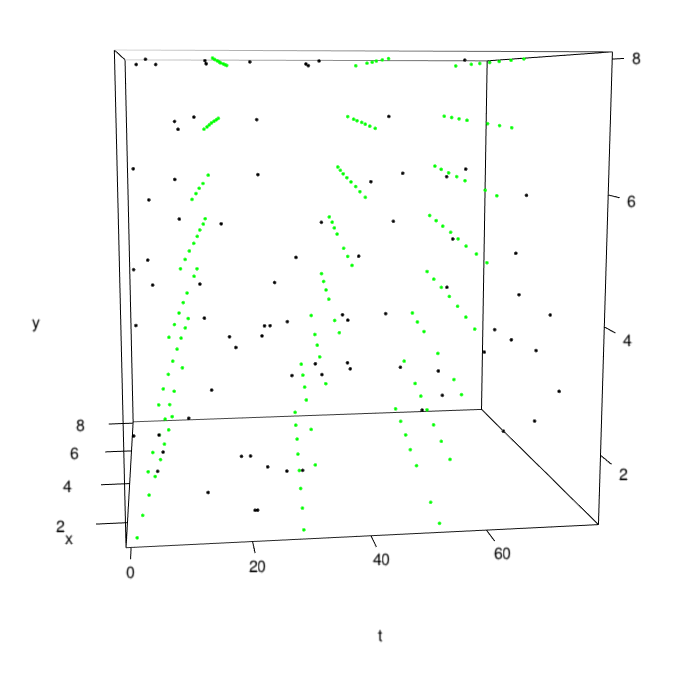}
	\caption[Simulated ground truth data of signal propagation in a grid
	of detector across time]
	{Simulated ground truth data of signal propagation in a grid
		of detector across time. The signals are propagating from $x=1$, $y=1$ and $t_0=1$. On the $x$-direction, the travelling time between adjacent electrodes follows a Normal distribution with $\mu_x = 1$, $\sigma_x = 0.1$. On the $y$-direction, the travelling time between adjacent electrodes follows a Normal distribution with $\mu_y = 2$, $\sigma_y = 0.2$. There is a probability $p=0.1$ that a spike is not detected. Multiples signals are propagated from the same starting location, with the successive starting time following an exponential distribution with $\lambda_s = \frac{1}{30}$. Noise are also generated with a uniform distribution across all electrode and time difference following exponential distribution with $\lambda_n = 1$.
	}
	\label{fig:houghdata}
\end{figure}

Figure \ref{fig:houghdata} gives an example of the
synthetic data with colored ground truth, where $t_{\max} = 80$, $\mu_x = 1$, $\sigma_x =
0.1$, $\mu_y = 2$, $\sigma_y = 0.2$, $p = 0.1$, $\lambda_s =
\frac{1}{30}$, $t_0 = 1$, $\lambda_n = 1$. The green dot represents
the position and time $(x,y,t)$ that a signal spike occurs, while
the black dot represents noise.

\paragraph{Performance of the randomized Hough transform}

\begin{figure}[!ht]
	\centering
	\includegraphics[width=\textwidth]{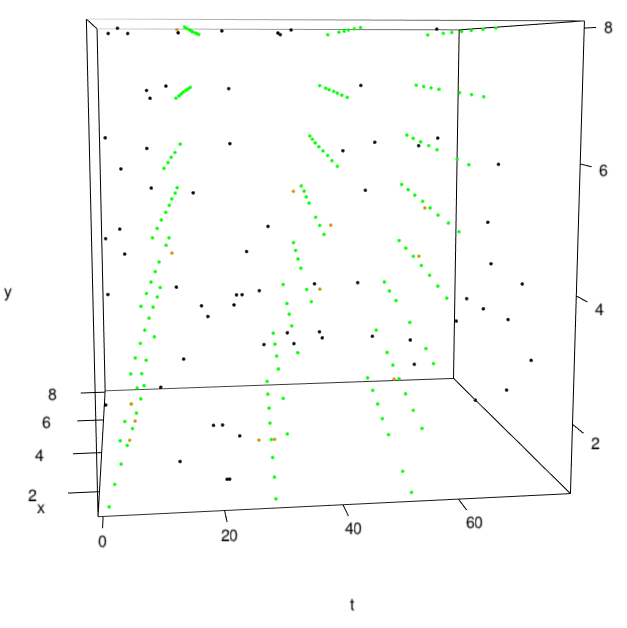}
	\caption[Hough transform of simulated data]
	{Hough transform of simulated data. The simulated data from Figure \ref{fig:houghdata} is separated into a number of waves using Hough transform. The result is plotted using the following color scheme: true position - green, true negative - black, false positive - yellow, false negative - red. All the signals are correctly classified. A few noises in close proximity to the signals are misclassified as signals.}
	\label{fig:houghresult}
\end{figure}

\begin{figure}[!ht]
	\centering
	\includegraphics[width=0.48\textwidth]{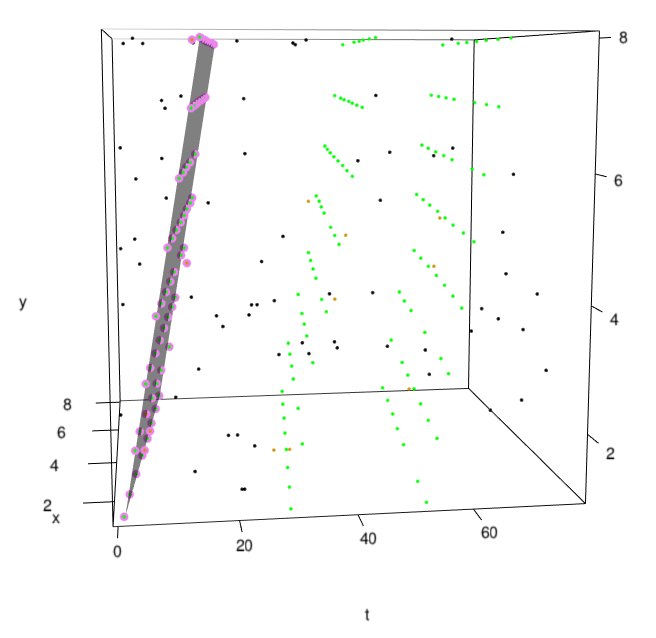}
	\includegraphics[width=0.48\textwidth]{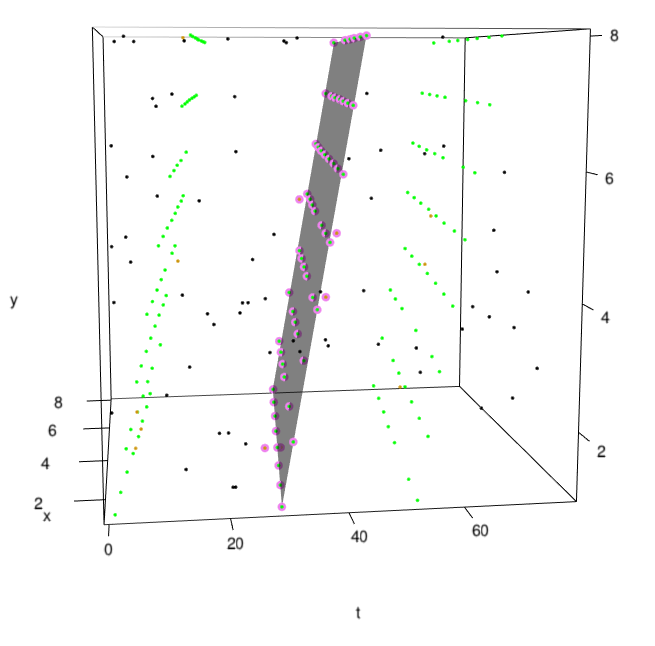}
	\includegraphics[width=0.48\textwidth]{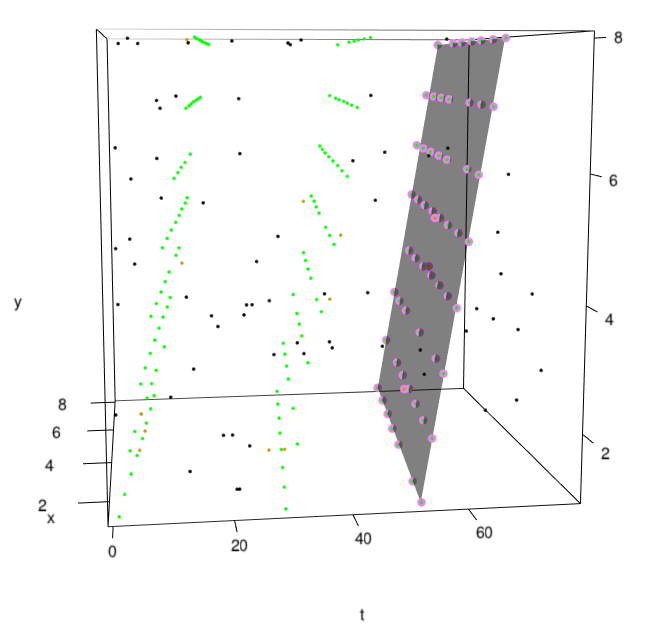}
	\caption[Planes found by Hough transform of simulated data]
	{Planes found by Hough transform of simulated data. The shaded
		area in three plots
		above represent each of the three planes found by the application of Hough
		transform in Figure \ref{fig:houghresult}.}
	\label{fig:houghresultplane}
\end{figure}

\begin{figure}[!ht]
	\centering
	\includegraphics[width=\textwidth]{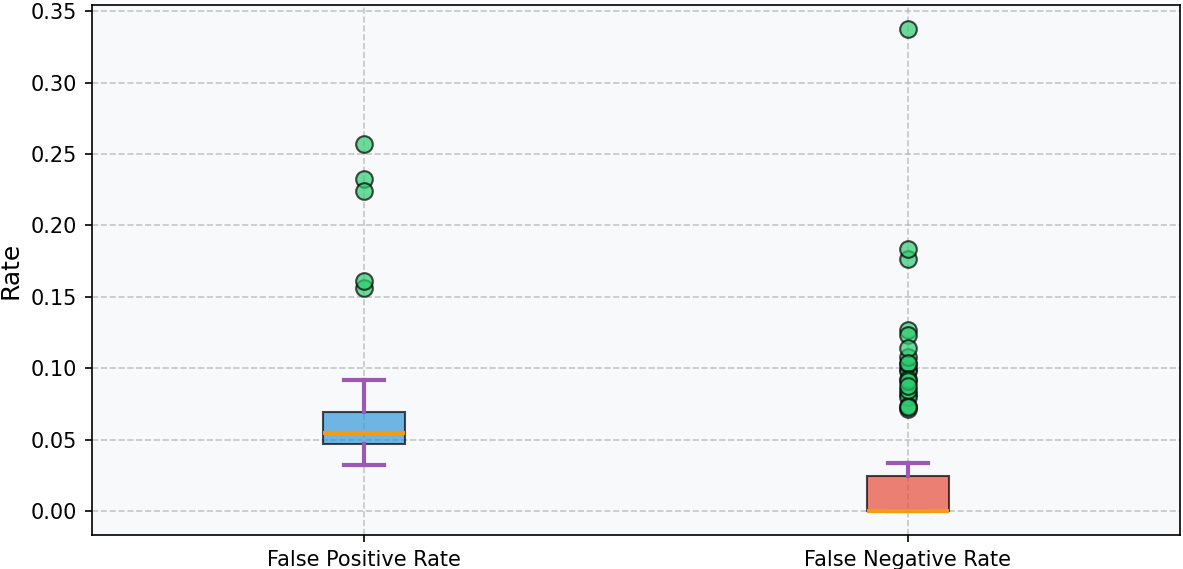}
	\caption[Boxplot of Performance Metrics]
	{Boxplot of Performance Metrics. It summarizes the distribution of the false positive rate (FPR) and false negative rate (FNR) for the evaluated algorithm. The FPR, represented by the blue box, has a mean of 0.065 (6.5\%) with a variance of 0.001333. The FNR, shown in red, demonstrates a lower mean of 0.0257 (2.57\%) but a higher variance of 0.002798. The green dots represent individual data points from the simulation runs.} 
	\label{fig:rateboxplot}
\end{figure}

Synthetic data from the linear wavefront model can be used to test the performance of our modified randomized Hough transform algorithm (Algorithm \ref{alg:randomhough}). Figure \ref{fig:houghresult} demonstrates the wave detection result. In the figure, the points in green represent the true signal spikes that are classified as signal spikes (i.e. true positive); the points in black are the true noise spikes that are classified as noises (i.e. true negative); yellow represents true noise spikes being misclassified as signal spikes (i.e. false positive), and red color shows the true signal spikes being misclassified as noises (i.e. false negative). The performance of our algorithm is quantified by the number of spikes that are false positive and false negative. From Figure \ref{fig:houghresult} we can see that there is zero false negative (i.e. no true signal spike is classified as noise) and there are no more than five false positives for each detected plane (i.e. only a few noises that are extremely close to the signal plane are misclassified as signal spikes). The planes found by the Hough transform are illustrated in Figure \ref{fig:houghresultplane}. The simulation is repeated 100 times with different random seeds. The false positive rate (number of false positive divided by the number of spikes) has a mean of 6.5\% and a standard deviation of 2.6\%. The false negative rate (number of false negative divided by number of spikes) has a mean of 3.6\% and a standard deviation of 5.3\%. Here we present the box plot of the performance of the result \ref{fig:rateboxplot}. In summary, the randomized Hough transform performs quite well in signal-wave detection.

\subsubsection{Model fitting for signal propagating pattern}
\paragraph{Data Synthesis from circular wavefront model}
\label{sec:syntheticcircular}
To examine the performance of Algorithm \ref{alg:alternating}, we generate a set of signal data that have the form of a circular wavefront propagating pattern.

In order to visualize the complete shape of the propagating wave, a larger MEA grid is used for this simulation study, which is 96-by-96. In addition, the circular wavefront signal is simulated in the following ways:

\begin{itemize}
	\item The wave source is at any point in Euclidean space, either inside or outside the MEA grid. Suppose that its coordinates are $(x_0, y_0)$, with $x$ denoting the row and $y$ denoting the column.
	\item The duration of the whole simulation study is $t_{max}$. The MEA stops recording signals when $t_{max}$ is reached.
	\item The signal spike is propagating in all directions simultaneously with an identical speed $v$.
	\item There is an independent measurement error $e$ of the recording time for each electrode, which follows a normal distribution with mean $0$ and variance $\sigma^2$.
	\item For each electrode, there is a probability $p$ that the signal spike cannot be detected.
	\item The wave is excited at time $t_0$.
	\item The MEA system also records noise spikes, the locations of which are randomly selected from all electrodes on the MEA grid. The time gap between two noise spikes follows an exponential distribution with mean $\lambda_n$.
\end{itemize}

\begin{figure}
	\centering
	\includegraphics[width=\textwidth]{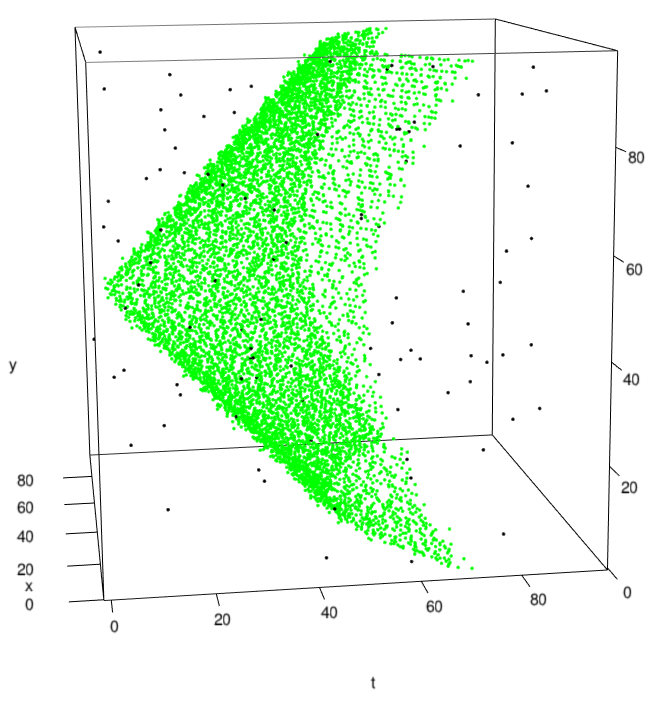}
	\caption[Simulated circular wavefront data]
	{Simulation of signal propagation forming a circular wavefront. The wave is propagated from $x_0 = 48$, $y_0 = 48$ at time $t_0 = 2$ and travelling at a speed of $v=1$. The recording time of each spike has a measurement error of $e$ that follows a normal distribution of mean 0 and variance $\sigma^2 = 1$. Signals have a missing probability of $p=0.1$. There are also noise spikes which are distributed uniformly in space and arrive in time difference with an exponential distribution of $\lambda_n = 1$. The green dots represents recording of signals and the black dots represent noises.}
	\label{fig:circulardata}
\end{figure}

An example of synthetic data is shown in Figure \ref{fig:circulardata} with
$x_0 = 48$, $y_0 = 48$, $t_{\max} = 96$, $v=1$, $\sigma = 1$, $p=0.1$, $t_0 = 2$, and $\lambda_n = 1$.

\paragraph{Performance of the model fitting algorithm}
With the synthetic circular wavefront model data, we can apply the alternating variable method to estimate the unknown variables. For a detailed explanation of the algorithm, see Section \ref{sec:gridoptim}.

\begin{figure}[H]
	\centering \includegraphics[width=0.48\textwidth]{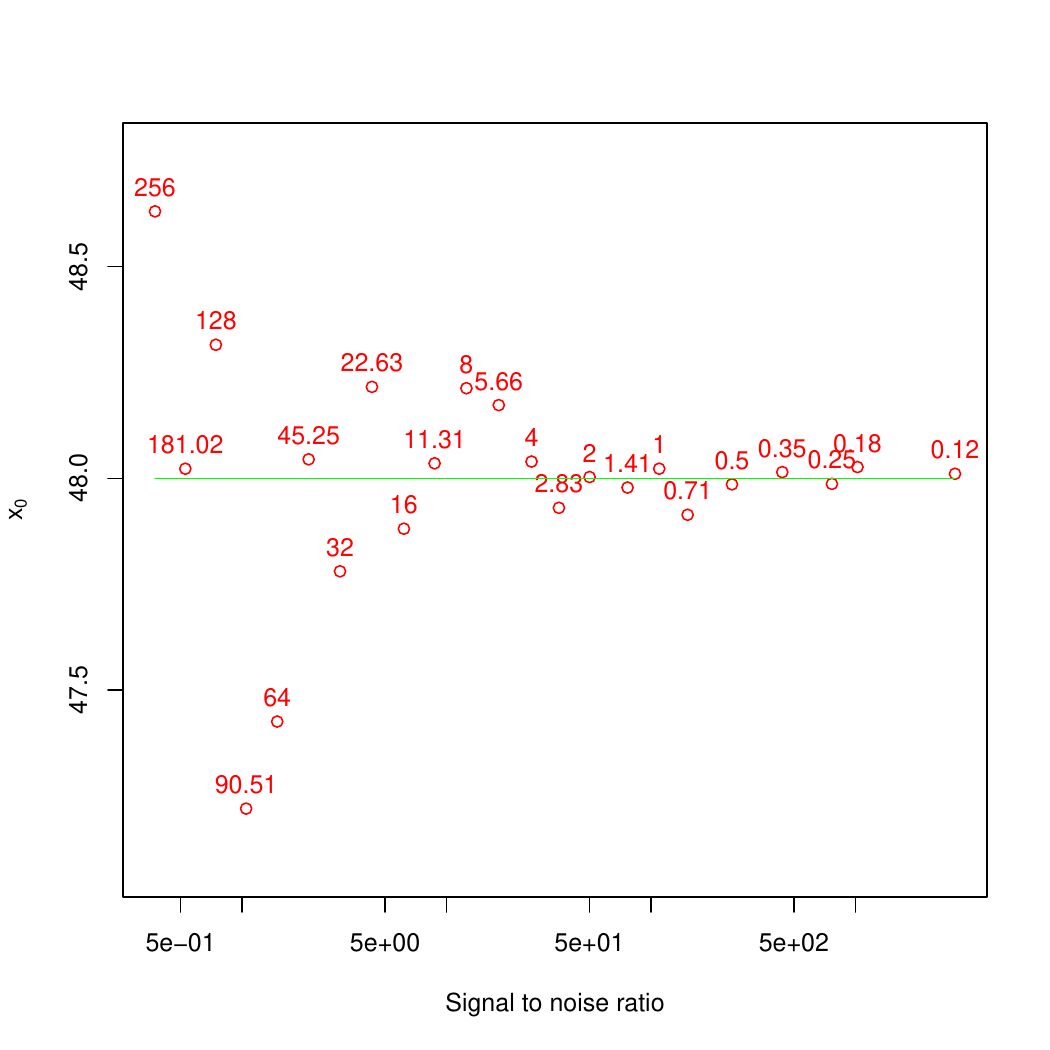}
	\includegraphics[width=0.48\textwidth]{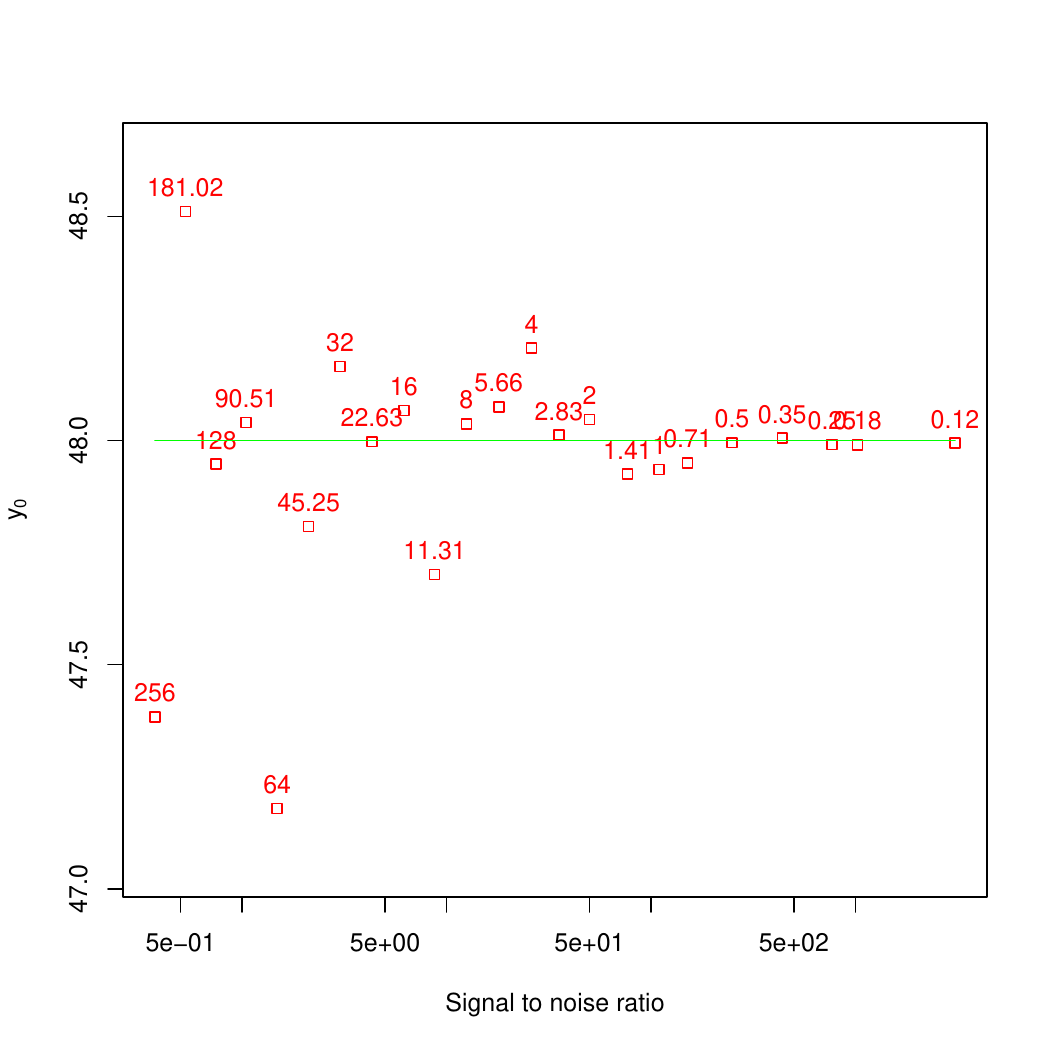}
	\includegraphics[width=0.48\textwidth]{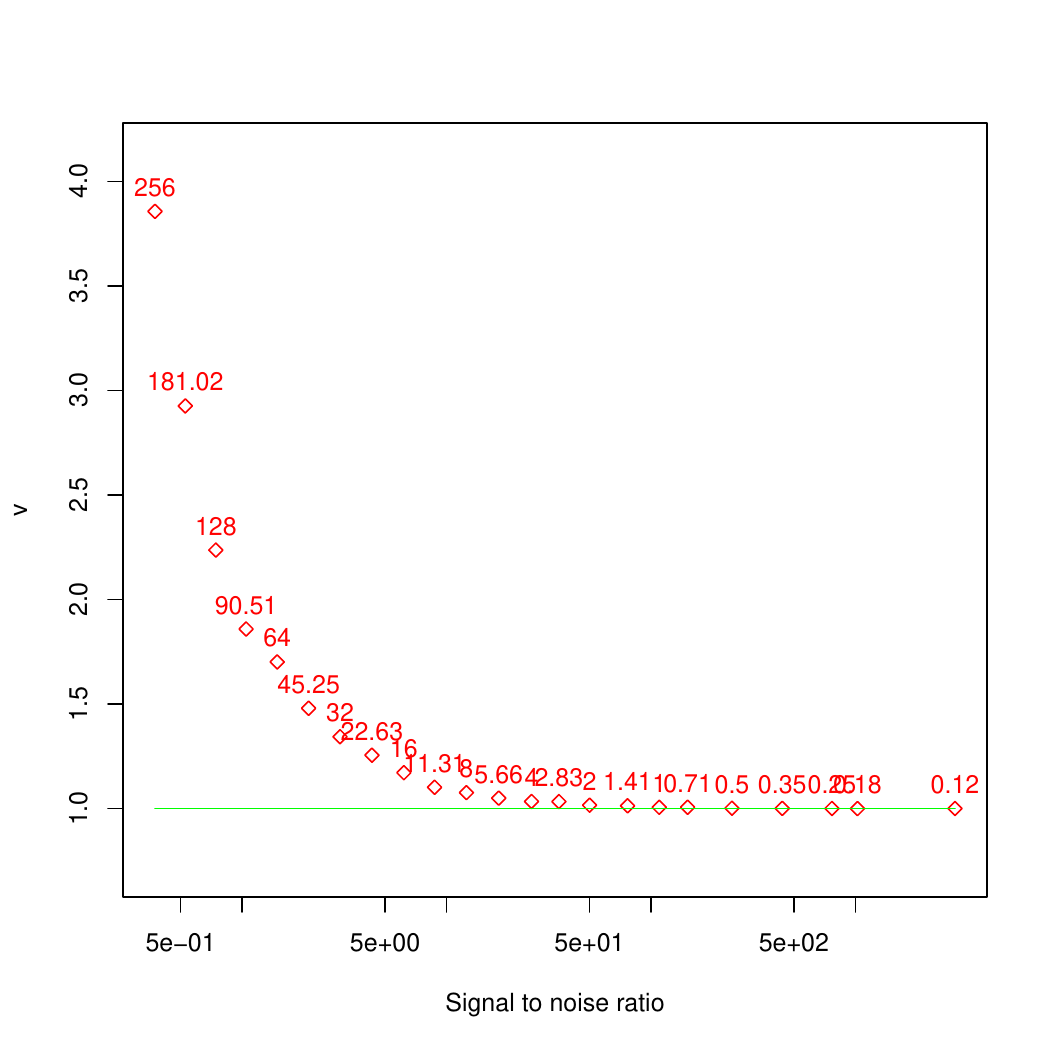}
	\includegraphics[width=0.48\textwidth]{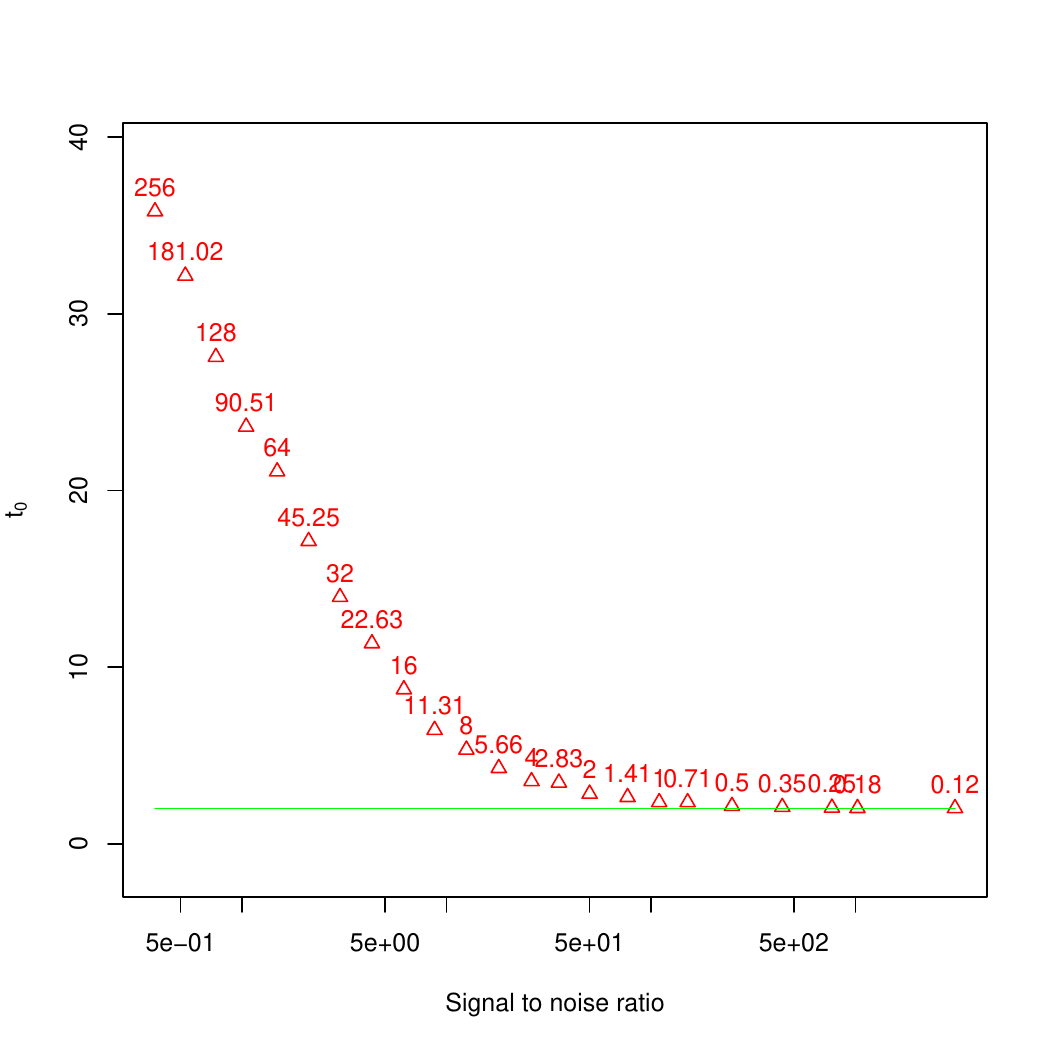}
	\caption[Result of fitting the circular model under different
	SNR]
	{Result of fitting the circular model under different signal-to-noise
		ratio (SNR). The above four plots shows the result of applying
		Algorithm \ref{alg:alternating} to the data simulated using the
		method described in Section \ref{sec:syntheticcircular}. We
		emphasize on the effect of
		noise frequency on the performance of estimation,
		so we set the signal missing probability and variance
		of measurement error to some very small values.
		In these plots, $v = 1$, $x_0=48$, $y_0=48$, $t_0=2$, $p=0$ and
		$\sigma=10^{-6}$.  In all of the above plots, $\lambda_n$, which
		governs the frequency of noise, varies from 256 to 0.12 (as
		indicated by the red label above each plot point). This corresponds
		to SNR from 0.37 to 3072. The true value of the parameter ($x_0$,
		$y_0$, $v$, $t_0$) is indicated by the green line on the plots.
		The upper left plot shows the value of $x_0$ fitted by our
		algorithm. When SNR is small, the variance in the estimate of
		$x_0$ is much larger. As the SNR becomes larger, our estimation
		of $x_0$ becomes more accurate. The same trend also applies to
		the upper right plot, which should the value of $y_0$. The lower
		left plot shows the value $v$ estimated by our algorithm. When
		the SNR is small, our estimate deviates significantly from the
		true value. When the SNR is big, our algorithm accurately found
		the value of $v$. The same trend also applies to the estimation
		of $t_0$, which is shown on the lower right plot.
	} \label{fig:change_lambda}
\end{figure}

\begin{figure}[H]
	\centering
	\includegraphics[width=0.48\textwidth]{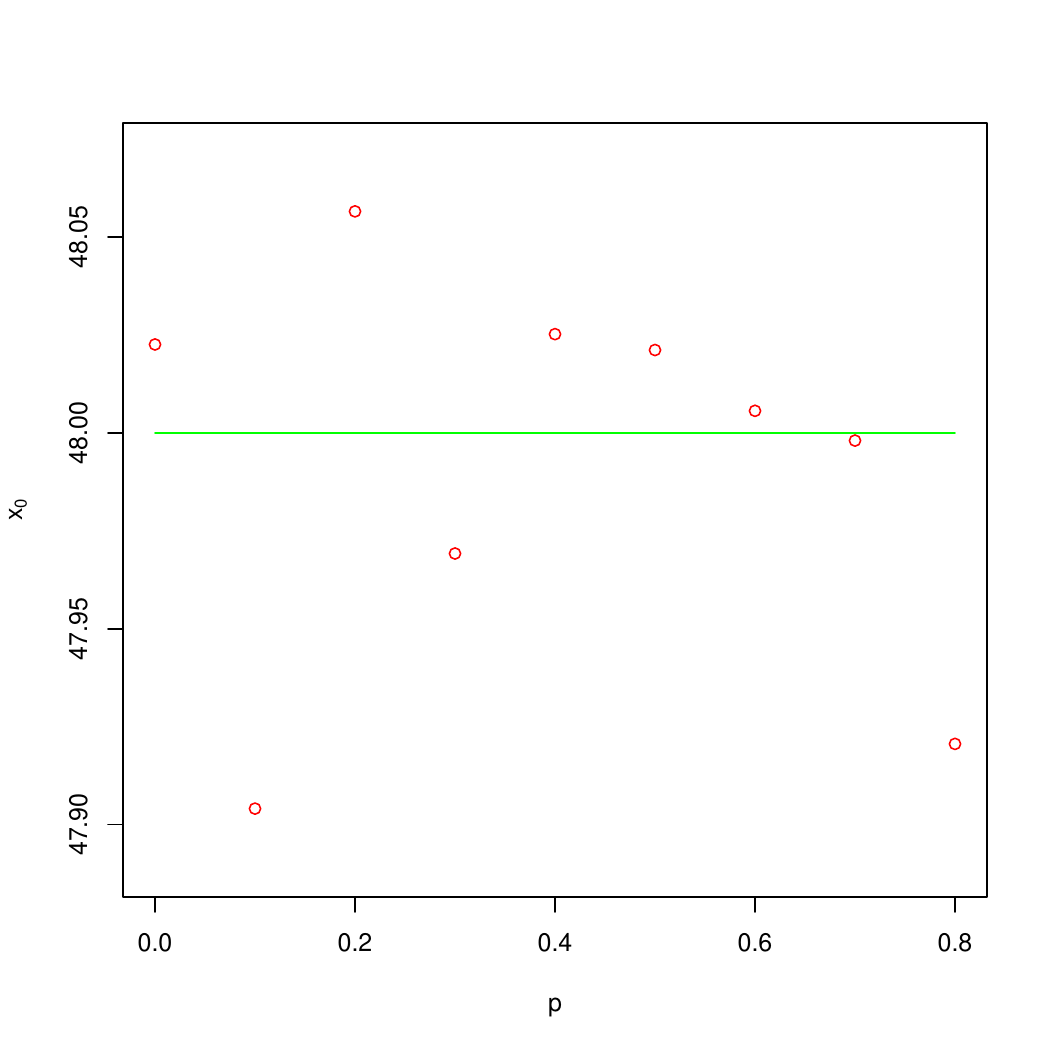}
	\includegraphics[width=0.48\textwidth]{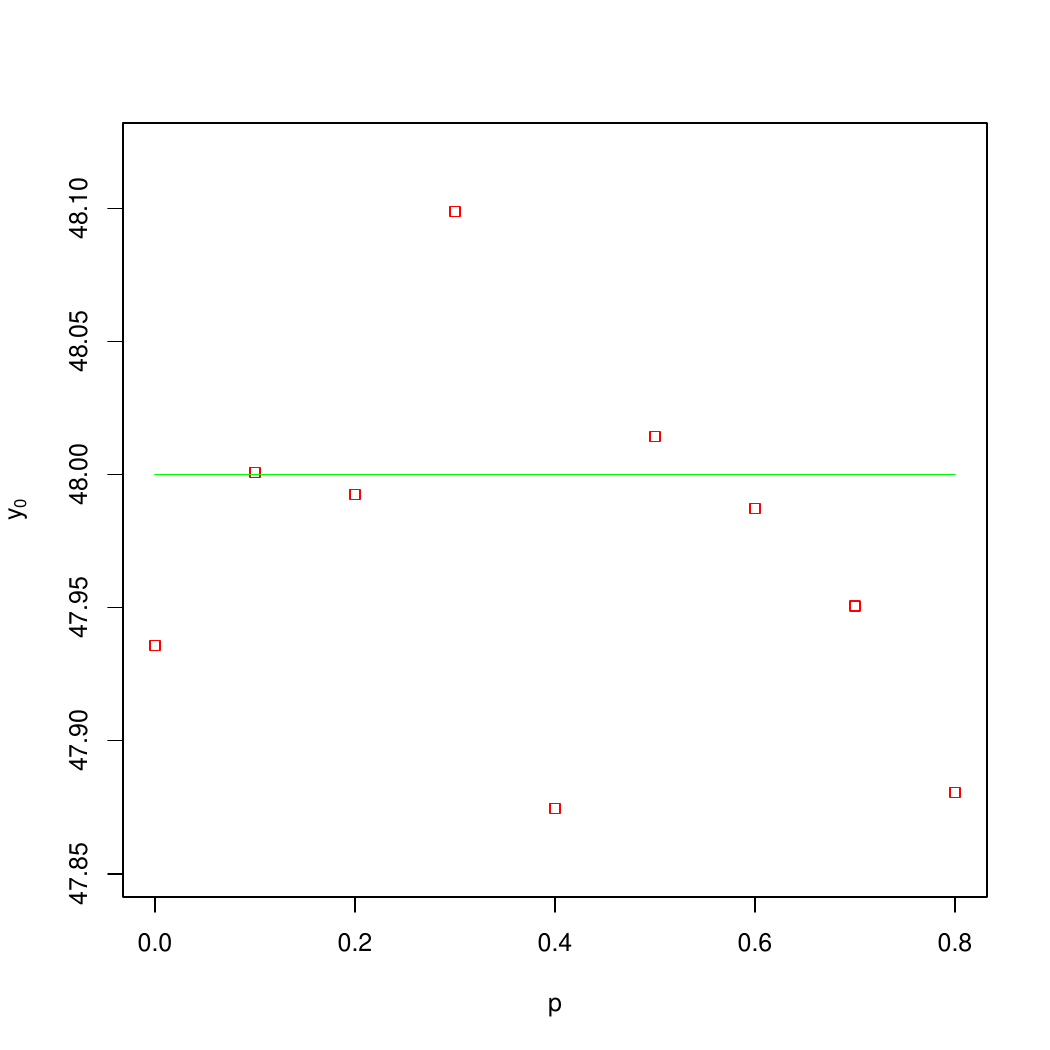}
	\includegraphics[width=0.48\textwidth]{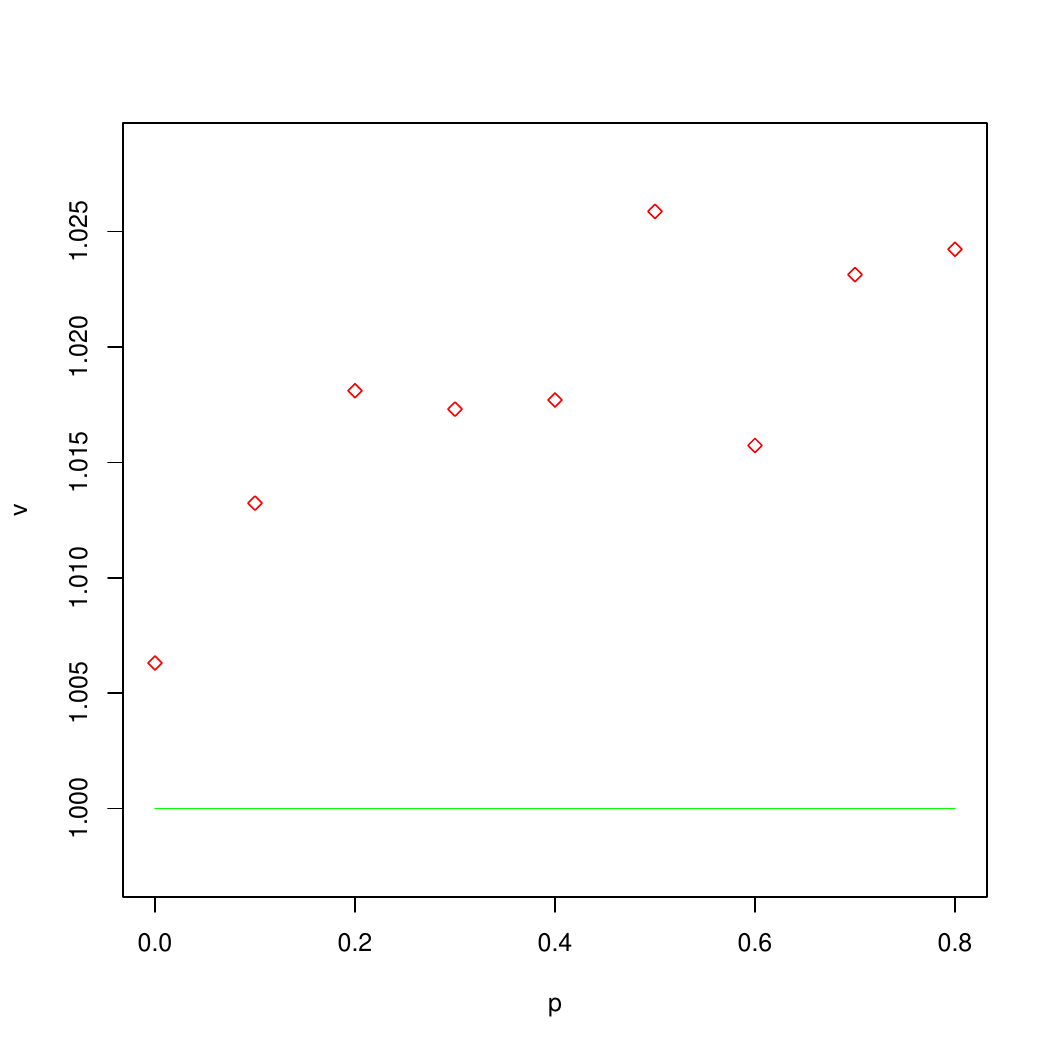}
	\includegraphics[width=0.48\textwidth]{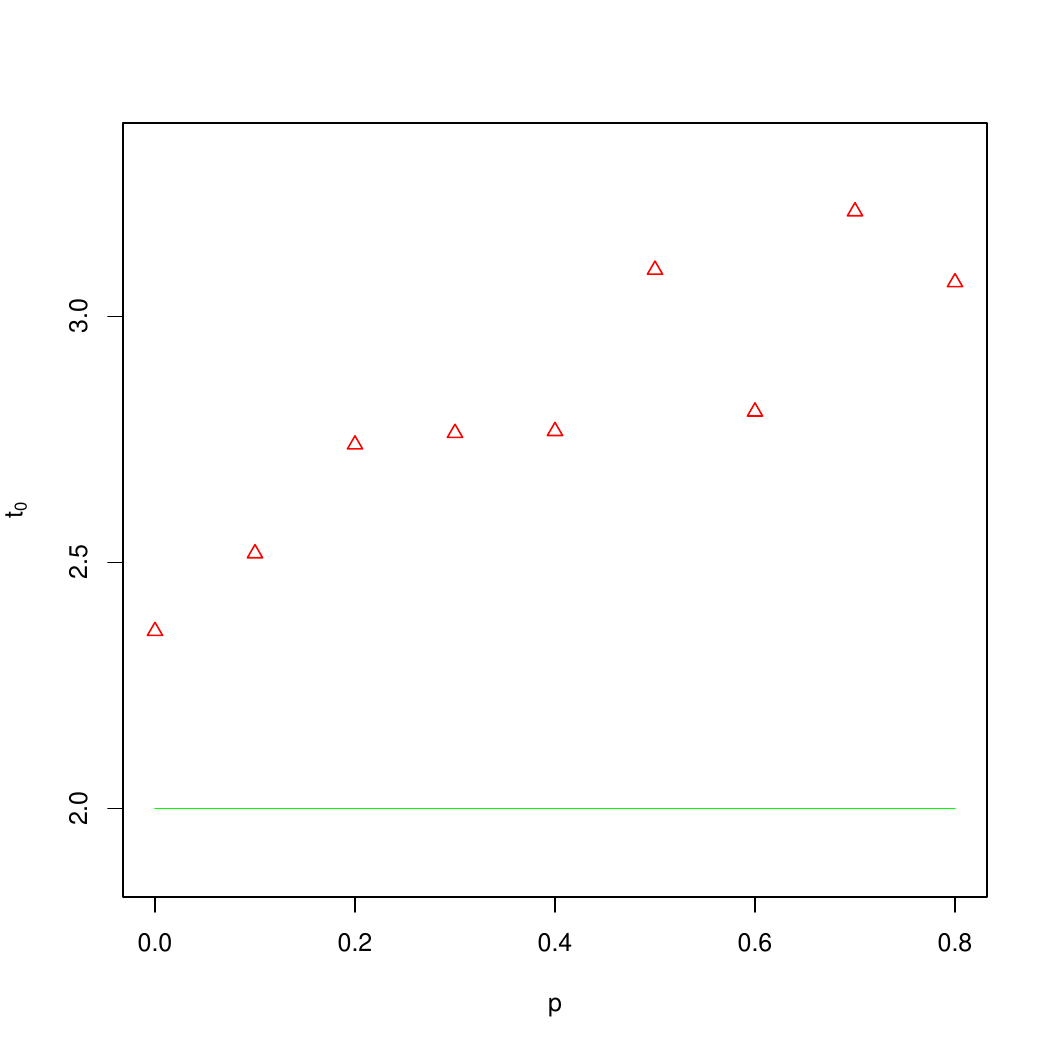}
	\caption[Result of fitting the circular model under different missing probability]
	{Result of fitting the circular model under different missing probability (p). The above four plots shows the result of applying Algorithm \ref{alg:alternating} to the data simulated using the method described in Section \ref{sec:syntheticcircular}. We emphasize on the effect of signal missing probability on the performance of estimation, so we fix the value of noise frequency and set the variance of measurement error to a very small value. In these plots, $v = 1$, $x_0=48$, $y_0=48$, $t_0=2$, $\lambda_n=1$ and $\sigma=10^{-6}$. In all of the above plots, $p$, which governs the signal missing probability, varies from 0 to 0.8. The true value of the parameter ($x_0$, $y_0$, $v$, $t_0$) is indicated by the green line on the plots.	The upper left plot shows the value of $x_0$ fitted by our algorithm, and the upper right plot shows the value of estimated $y_0$. As the signal missing probability increases from 0 to 0.8,our algorithm locates the wave source quite accurately. The estimated $x_0$ ranges from 47.90 to 48.05, and the estimated $y_0$ ranges from 47.87 to 48.10. The lower left plot shows the value $v$ estimated by our algorithm. When $p$ is small, the deviation of the estimated $v$ from the true value is much smaller. As the $p$ becomes larger, our estimation of $v$ becomes less accurate.The same trend also applies to the estimation of $t_0$, which is shown on the lower right plot.}
	\label{fig:change_p}
\end{figure}

In this simulation study, we generate data according to different noise frequencies and different signal missing probability.  We then apply Algorithm \ref{alg:alternating} to the simulated data. The estimated parameters ($x_0$, $y_0$, $v$, $t_0$) that characterize the signal propagation pattern under different signal-to-noise ratio and signal missing probability are shown in Figures \ref{fig:change_lambda} and \ref{fig:change_p}.

Figure \ref{fig:change_lambda} presents the result of fitting the circular model at different noise frequencies ($\lambda_n$), and thus with a different signal-to-noise ratio (SNR), which is calculated by the number of signals over the number of noises. The red points represent the estimated parameter values, while the green lines indicate the true values. We can see that as SNR increases, the fitted values of ($x_0$, $y_0$, $v$, $t_0$) become more accurate. The same trend also applies to Figure \ref{fig:change_p}, which gives the values of four estimated parameters under different signal missing probability ($p$).  The smaller $p$, the better the estimation. Although the missing probability reaches a relatively high value, our algorithm performs quite well. A similar analysis is performed on the standard deviation $\sigma$ of the measurement error. As $\sigma$ decreases, the estimation error also decreases accordingly.

\subsection{Real Data Study}
In this chapter, we apply the three-step process proposed in Section \ref{fig:Process_Flow} to a real data set. In this study, we used MEA to detect extracellular electrical potential signals and calculate their speed and location for further pharmacological, biological, and analytical studies.

\subsubsection{Data Preparation}

\begin{figure}[H]
	\centering
	\includegraphics[width=0.8\textwidth]{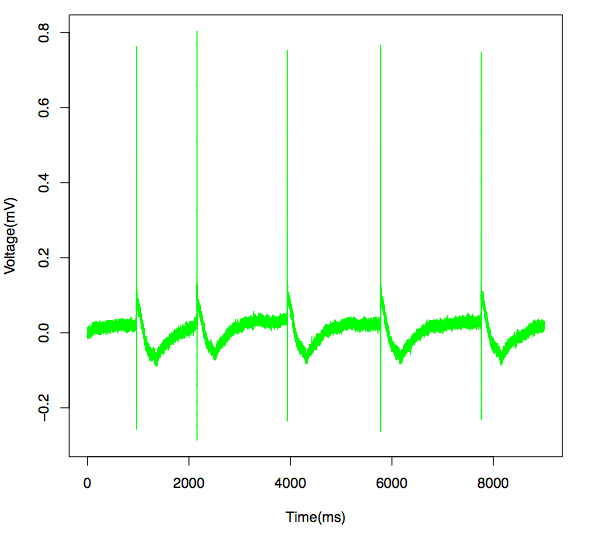}
	\caption[Traces of the extracellular field potentials of the cardiomyocytes from one electrode]
	{Traces of the extracellular field potentials of the cardiomyocytes detected by one electrode of the 8\texttimes 8 MEA. The whole recording period lasts 9 seconds.}
	\label{fig:onechannel}
\end{figure}

Cultured heart cells have been used in many studies of cardiac pathophysiology. In our investigation, cardiac myocytes were obtained from Sprague-Dawley rats on embryonic day 16-19, and were cultured in MEAs according to a previously published protocol \cite{bib31}. The MEA recorded 9 responses under control conditions, that is, there is no presence of drugs and the culture medium was under normoxic conditions, containing 92.5\% to 95.2\% dissolved $\textrm{O}_2$. The MEA chips that we used for the extracellular recordings are the 64-channel planar gold MEAs. The extracellular field potentials of cardiomyocytes from one channel are presented in Figure \ref{fig:onechannel}.

After recording the signal data, for further processing, signal spikes should be exacted. Usually, for each detection, noise is always unavoidable, so we have to reduce the noise from the data. Since what we received are frequency signals, we can use a filter to deal with them, and here, after verification, it has been found that the 4th order Butterworth high-pass filter provides the best noise reduction effect. Subsequently, the 99.95th percentile of the filtered signal is taken as the threshold. Only when a signal peak exceeds this threshold will it be marked as a spike for subsequence computation.

In our study, we used an 8*8 MEA matrix, which means that we will receive different signal data from 64 channels. Therefore, in a unit of time, the positional parameters of the spike data will reach three dimensions, and that is how the parameter set $\{(x_i, y_i, t_i)\}_{i=1}^n$ arrives.

\subsubsection{Identify multiple propagating waves}
\begin{figure}[!ht]
	\centering
	\includegraphics[width=0.8\textwidth]{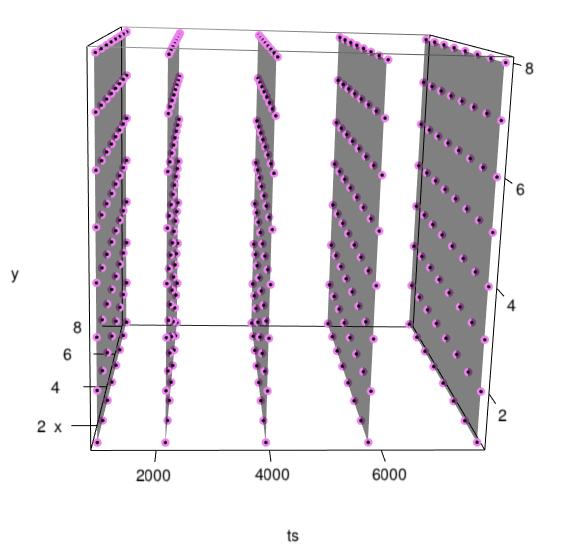}
	\caption[Identify the multiple propagating waves using Hough transform]
	{Hough transform of real data. The signal spikes (purple points) extracted by the Butterworth approach
		are separated into five planes (grey planes) using the randomized Hough
		transform (Algorithm \ref{alg:randomhough}). Each plane represents a propagating wave.
		The number of planes (i.e. 5) is consistent with the true number of action potentials shown in Figure \ref{fig:onechannel}.}
	\label{fig:houghreal}
\end{figure}

After extracting the spike activation times for each electrode, we convert the channel number into the Cartesian coordinate $(x,y)$, where $x$ and $y$ denote the row and column number of the electrode on the MEA grid respectively. Hence, we obtain a set of tuples $\{(x_i, y_i, t_i)\}_{i=1}^n$ representing the spikes in the three-dimensional space $\{(x, y, t)\}$. In order to derive the beat frequency of the untreated cardiomyocytes under normoxic condition, Randomize Hough transform (Algorithm \ref{alg:randomhough}) will be performed. Figure \ref{fig:houghreal} shows the result of the propagating waves detection. We can see that the number of detected planes is five, which is the same as the true number of signal waves presented in Figure \ref{fig:onechannel}.

Since the whole experiment is about processing cultured heart cells' bio-electricity signal, and usually, such signals are single-source, therefore, for better fitting the signal source, we can slightly adjust the original algorithm:

1. Master Normal Vector Selection: The normal vector $\vec{n}$ of the plane containing the most points is designated as the master direction of propagation. This embeds the prior biological knowledge that a dominant wavefront should exist.

2. $5^{\circ}$ Tolerance Constraint: The normal vector $\vec{p}$ of every other detected plane is compared to $\vec{n}$. Any plane for which the angle between $\vec{p}$ and $\vec{n}$ exceeds $5^{\circ}$ is discarded. This stringent angular tolerance ensures that all accepted waves originate from a consistent, single source direction, effectively filtering out aberrant detections caused by noise or non-physiological propagation.

The altered algorithm~\ref{alg:randomhoughalt} is shown as follows.

\begin{algorithm}
	\caption{Randomize Hough transform for bio-electricity wave detection}
	\label{alg:randomhoughalt}
	\begin{algorithmic}
		\State Given a set of spikes $\{(x_i, y_i, t_i)\}_{i=1}^n$
		\While{}
		\State Run randomized Hough transform until a bucket gets 8 votes
		\State Check every remaining point to see if it is in the plane
		spanned by this bucket
		\If{Number of points on the plane $> \frac{2}{3}$ (Number of detectors)}
		\State Mark the points on this plane as a cluster and remove all
		these points
		\Else
		\State Exit while loop
		\EndIf
		\EndWhile
		\State Set $\vec{n}$ to be the normal of the plane with most points
		\For{each plane $\mathcal{P}$ found in the above while loop}
		\State Set $\vec{p}$ to be the normal of $\mathcal{P}$
		\If{angle between $\vec{n}$ and $\vec{p} > 5^\circ$}
		\State Remove plane $\mathcal{P}$ and mark all the points on it as
		unclassified
		\EndIf
		\EndFor
		\For{Each point $P$ that's not marked to any plane}
		\If{$P$ is sufficiently close to a plane found above}
		\State Add $P$ to that plane.
		\EndIf
		\EndFor{}
		\For{Each point $P$ that's not marked to any plane}
		\State Check if every other point belongs to a plane with normal $\vec{n}$
		and pass through $P$.
		\If{Number of points matching the above criteria $>
			\frac{1}{2}$ (Number of detectors)}
		\State Mark all of them as a plane
		\EndIf
		\EndFor
		\State The unclassified points will be regarded as noise.
	\end{algorithmic}
\end{algorithm}

\subsubsection{Model fitting}
For each plane in Figure \ref{fig:houghreal}, we fit both the circular and linear wavefront model to estimate the unknown parameters that describe the propagating pattern of spike activities.

\paragraph{Fitting the circular wavefront model}
We apply Algorithm \ref{alg:alternating} for fitting the circular wavefront model. $(x_0, y_0)$ represents the origin of the excitation. $t_0$ represents the time of excitation, and $v$ is the speed of wave propagation. The circular model is particularly useful when the wave source is inside, or close to the MEA grid. For our experimental data, the MEA grid lies in the region $[1,8] \times [1,8]$. When searching for the origin or excitation, we pick a search space $[-50, 50] \times [-50, 50]$ that is big enough to cover the neighbourhood of the MEA grid. If the origin of the excitation found by circular model lies on the boundary of our search space, that would imply the wave is originated from far away, and it's more appropriate to use the linear wavefront model in such cases.

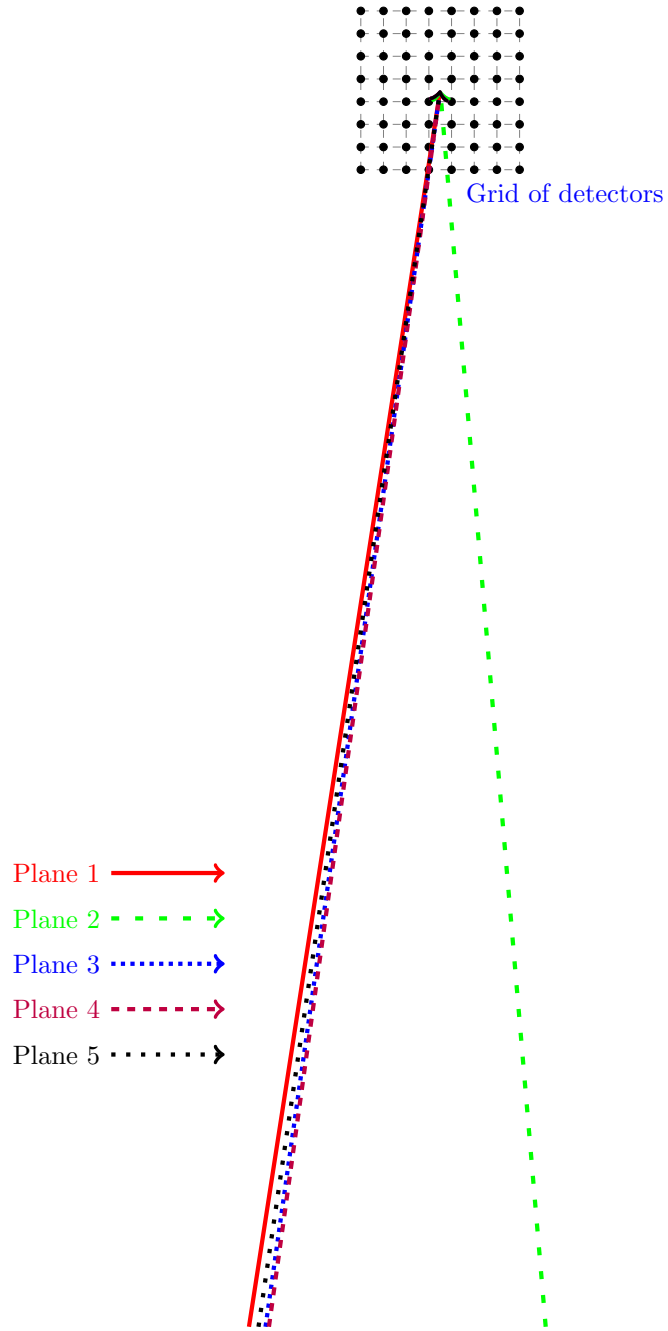
\begin{figure}[H]
	\centering
	\scalebox{1}{
		\begin{tikzpicture}[scale=0.3]
			\draw[style=help lines,dashed] (1,1) grid[step=1cm] (8,8);
			\foreach \x in {1,2,...,8}{
				\foreach \y in {1,2,...,8}{
					\node[draw,circle,inner sep=1pt,fill] at (\x,\y) {};
				}
			}
			\coordinate (Centroid) at (4.5,4.5);
			\draw [ultra thick,red,->] (-3.93, -50)  -- (Centroid);
			\draw [ultra thick,loosely dashed,green,->] (9.15,  -50)  -- (Centroid);
			\draw [ultra thick,blue,dotted,->] (-3.20, -50) -- (Centroid);
			\draw [ultra thick,purple,dashed,->] (-3.05, -50) -- (Centroid);
			\draw [ultra thick,black,loosely dotted,->] (-3.51, -50)  -- (Centroid);
			
			\draw [ultra thick,red,->] (-10, -30) node [left] {Plane 1} -- (-5,-30);
			\draw [ultra thick,loosely dashed,green,->]
			(-10, -32) node [left] {Plane 2} -- (-5, -32);
			\draw [ultra thick,blue,dotted,->]
			(-10, -34) node [left] {Plane 3} -- (-5, -34);
			\draw [ultra thick,purple,dashed,->]
			(-10, -36) node [left] {Plane 4} -- (-5, -36);
			\draw [ultra thick,black,loosely dotted,->]
			(-10, -38) node [left] {Plane 5} -- (-5, -38);
			
			\draw [blue]
			(10,0) node {Grid of detectors};
		\end{tikzpicture}
	}
	\caption[Routes of signal propagation using circular wavefront model]{
		Routes of signal propagation using circular wavefront model. The five arrows in the above graph points from the wave source found on each plane using the circular wave model to the centroid of the MEA. Thus, they represent the direction that each wave is propagating. We can see that plane 1, 3, 4, 5 are propagating in the same direction, while plane 2 comes from a slightly different direction.}
	\label{fig:circulardirection}
\end{figure}

Table \ref{tab:circularreal} presents the result of fitting the circular wavefront model to each propagating wave. We can see that for all 5 waves, $(x_0, y_0)$ found by our Algorithm \ref{alg:alternating} lies on the boundary of $[-50, 50] \times [-50, 50]$, with $y_0$ always being -50. Thus, we conclude that the wave comes from outside our search space. The wave source found here also provide us with a rough estimation of the direction that our wave is propagating. Since we are fitting a model of wave passing through our MEA grid, the direction from $(x_0, y_0)$ to the centroid of the MEA grid $(4.5, 4.5)$ gives us the direction of each wave. This is visualized in Figure \ref{fig:circulardirection}. Also, all 5 waves have similar estimated $v$, which conforms to our expectation that under the same condition multiple signal waves are propagating on the MEA with similar speed.

\begin{table}[!ht]
	\caption[Result of fitting the circular wavefront model]{Result of fitting the circular wavefront model.}
	\label{tab:circularreal}
	\centering
	\begin{tabular}{*{5}{c}}
		Plane   & $x_0$ & $y_0$ & $v$ & $t_0$  \\
		\hline
		1       & -3.93 & -50.00 & 0.44 & 849.75 \\
		2       & 9.15 & -50.00 & 0.49 & 2054.29 \\
		3       & -3.20 & -50.00 & 0.46 & 3828.32 \\
		4       & -3.05 & -50.00 & 0.46 & 5667.08 \\
		5       & -3.51 & -50.00 & 0.45 & 7647.11 \\
	\end{tabular}
\end{table}

\paragraph{Fitting the linear wavefront model}

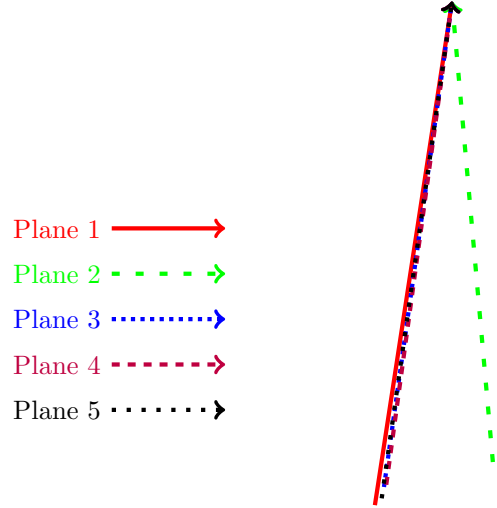
\begin{figure}[H]
	\centering
	\scalebox{1}{
		\begin{tikzpicture}[scale=3]
			\coordinate (Centroid) at (0,0);
			\draw [ultra thick,red,->] (-0.34, -2.22)  -- (Centroid);
			\draw [ultra thick,loosely dashed,green,->] (0.18, -2.03)  -- (Centroid);
			\draw [ultra thick,blue,dotted,->] (-0.30, -2.14) -- (Centroid);
			\draw [ultra thick,purple,dashed,->] (-0.29, -2.13) -- (Centroid);
			\draw [ultra thick,black,loosely dotted,->] (-0.31, -2.19)  -- (Centroid);
			
			\draw [ultra thick,red,->] (-1.5, -1) node [left] {Plane 1} -- (-1,-1);
			\draw [ultra thick,loosely dashed,green,->]
			(-1.5, -1.2) node [left] {Plane 2} -- (-1, -1.2);
			\draw [ultra thick,blue,dotted,->]
			(-1.5, -1.4) node [left] {Plane 3} -- (-1, -1.4);
			\draw [ultra thick,purple,dashed,->]
			(-1.5, -1.6) node [left] {Plane 4} -- (-1, -1.6);
			\draw [ultra thick,black,loosely dotted,->]
			(-1.5, -1.8) node [left] {Plane 5} -- (-1, -1.8);
			
		\end{tikzpicture}
	}
	\caption[Routes of signal propagation using linear wavefront model]{
		Routes of signal propagation using linear wavefront model. The five arrows above represents the direction $(\ta,\tb)$ found on the five planes. The directions closely match those from circular wavefront model in Figure \ref{fig:circulardirection}. Wave 1,3,4,5 goes in the same direction while wave 2 goes in a slightly different direction.
	}
	\label{fig:lineardirection}
\end{figure}

As mentioned in Section \ref{sec:linearfit}, we use the linear least squared method to estimate the parameters in the linear wavefront model.  In this model, the propagation of the wave is characterized by the direction $(\ta,\tb)$ and the speed $v$. The origin of the wave is assumed to be very far away.  Thus, the concepts of wave source and time of excitation do not apply here. The estimated parameters $\ta, \tb, v$ are shown in Table \ref{tab:linearreal}. The propagation direction represented by $(\ta, \tb)$ is visualized is Figure \ref{fig:lineardirection}.

\begin{table}[h]
	\caption[Result of fitting the linear wavefront model]{Result of fitting the linear wavefront model.}
	\label{tab:linearreal}
	\centering
	\begin{tabular}{*{4}{c}}
		Plane   & $\ta$ & $\tb$ & $v$  \\
		\hline
		1 &  0.34  & 2.22 & 0.44 \\
		2 & -0.18  & 2.03 & 0.49 \\
		3 &  0.30  & 2.14 & 0.46 \\
		4 &  0.29  & 2.13 & 0.46 \\
		5 &  0.31  & 2.19 & 0.45 \\
	\end{tabular}
\end{table}

\paragraph{Comparison of the two models}
Comparing Figure \ref{fig:circulardirection} and \ref{fig:lineardirection}, we can see that the direction of wave propagation found by the two methods closely match each other. All waves come from the south of our MEA grid. Comparing Table \ref{tab:circularreal} and \ref{tab:linearreal}, the wave propagation speed found by the two methods matches. The wave is travelling at a speed of 0.44-0.49 detectors/ms.

\newcommand{\textcircular}{\textrm{circular}}
\newcommand{\textlinear}{\textrm{linear}}
\newcommand{\textres}{\textrm{res}}
\newcommand{\texttot}{\textrm{tot}}
To compare how good each model fits our dataset, we define the following
coefficient of determination for the circular wavefront model
\begin{align*}
	R^2_{\textcircular}
	&= 1 - \frac{SS^\textcircular_\textres}{SS_\texttot}
	\\&= 1 - \frac{\closs}{\sum (t_i - \bar{t})^2}
	& \textrm{(from \eqref{eqn:circularloss}).}
\end{align*}
Similarly, for linear wavefront model, we have
\begin{align*}
	R^2_{\textlinear}
	&= 1 - \frac{SS^\textlinear_\textres}{SS_\texttot}
	\\&= 1 - \frac{\lloss}{\sum (t_i - \bar{t})^2}
	& \textrm{(from \eqref{eqn:linearloss}).}
\end{align*}

The coefficients of determination for each of the five planes fitted by our two methods are summarized in Table \ref{tab:Rsquare}. We can see that for each of the five planes, the circular wavefront model gives a slightly better fit than linear wavefront model.

\begin{table}[h]
	\caption[Coefficients of determination]
	{Coefficients of determination for circular and plane wavefront model}
	\label{tab:Rsquare}
	\centering
	\begin{tabular}{*{3}{l}}
		Plane   & $R^2_{\textcircular}$ & $R^2_{\textlinear}$ \\
		\hline
		1 &  0.948072   &   0.9426517    \\
		2 &  0.9593685   &  0.9566555     \\
		3 &  0.9373304   &  0.9308701     \\
		4 &  0.944263   &   0.9384128    \\
		5 &  0.9405785   &  0.934337     \\
	\end{tabular}
\end{table}

\section{Conclusion}\label{sec4}

In this paper, we introduced a RHT based wave detection method with a detector array that is used. We first detect signal data, remove noise and extract spike of it, then integrate the spikes' coordinates with time to form three-dimensional data. After that, we used the Randomize Hough transform to confirm the wave planes. Finally, to fit the model, we minimize loss functions to get the wave source coordinates or its direction of propagation, with propagation time and velocity. Subsequently, we use a simulated study and a real data study to substantiate the practicability and precision of our proposed method.

In the process of handling spike data, instead of processing two-dimensional data as static and incoherent forms, we added time in the data set to form three-dimensional data as a dynamic form. The wave front naturally forms a plane in the x-y-t space and combined with the adaptation of RHT, the accuracy and flexibility of processing wave data to get required result will be enhanced. Besides, in order to further enhance the accuracy of bio-electricity signal detection like the case we mentioned in experiment section, we introduced an enhanced RHT method to detect wave source in single source point case.

By using wave plane data generated by RHT, we also able to get the orientation of the source of wave, and acquire wave's propagation velocity and initial time value, which can help us reconstruct the complete signal propagation path and significantly promoting subsequent related research. The steps described here achieve a closed-loop process of "detection-separation-fitting", and have demonstrated extremely low false detection rates in both simulation and experimental data, verifying the robustness and high accuracy of the method.

In addition, in the process of establishing the initial model, we assumed that all waves have uniform speed on the grid and propagate in the forward direction. The high accuracy of this model was confirmed in the real data experiment, but it can be adjusted to fit non-linear diffusion or irregular electrode spacing cases. In future research, by doing this, the precision and accuracy of the model can be improved.

In our study above, what we take into consideration are two cases: wave detection in single source, and wave detection in multiple sources, and to cope with these two cases the methods we used are slightly different. So if mixture situation is encountered (like multi-waves are detected from a single source in multiple sources case), thing can be difficult to deal with. For this, DBSCAN could be introduced to achieve clustering of multiple parallel waves, and then when dealing with a situation of multiple waves originating from a single source, these multiple parallel waves will converge and form a single source coordinate rather than multiple ones.

Meanwhile, though efficiency of RHT is already much better than normal HT, the burden of calculation still remains significant, especially when number of detector units in the detector array becomes too much, for example, if 64*64 MEA is used a large amount of computation will be consumed. In this case, by using GPU parallelism, the computing process can be accelerated. In addition to the efficiency, as we can see in the experiment part, the algorithm applied in single-source detection cases and multi-source cases is different since a restrictive measure for improving robustness is adopted in the single-source case. For further improvement, the DBSCAN clustering induction can be applied in the original Algorithm~\ref{alg:randomhough} to handle both cases (single-source and multi-source) at the same time.

\section*{Availability}
The code and data of this work are available on GitHub at \text{https://github.com/Fansese/wave-hough-detect}.

\section*{Acknowledgment}
We sincerely thank Professor John A Rudd for introducing the gastric microelectrode array data to us, which motivated this research.







\bibliography{sn-bibliography}



\end{document}